# Ways to Reduce Cost of Living: A Case Study among Low Income Household in Kubang Pasu, Kedah, Malaysia

*Siti Hadijah Che Mat\*, School of Economics, Finance and Banking, Universiti Utara Malaysia, Sintok, Kedahh, Malaysia.*

*Munzarina Ahmad Samidi, School of Economics, Finance and Banking, Universiti Utara Malaysia, Sintok, Kedahh, Malaysia.*

*Mukaramah Harun, School of Economics, Finance and Banking, Universiti Utara Malaysia, Sintok, Kedahh, Malaysia.*

*Wan Roshidah Fadzim, School of Economics, Finance and Banking, Universiti Utara Malaysia, Sintok, Kedahh, Malaysia.*

*Mohd Saifoul Zamzuri Noor, School of Economics, Finance and Banking, Universiti Utara Malaysia, Sintok, Kedahh, Malaysia.*

**Abstract**--- This study was conducted to examine and understand the spending behavior of low income households (B40), namely households with income of RM3800 and below. The study focused on the area Kubang Pasu District, Kedah. The data was collected from face to face interview based on structured questionnaire to respondents consisting of 124 head of household, randomly selected in the Kubang Pasu district. The findings reveal that the majority of these low-income household spend bigger percentage of their income on food items. Head of household work mostly in agricultural sector, forestry and fisheries, which accounted for 27%, and most of them have lower than RM2000 of monthly income. Respondents also stated that they have to engage in part time work and were forced to save money to cope with the rising cost of living. Various suggestions given by households to cope with the rising livingcosts which include the elimination of GST, financial aid by the government, and price control for goods and fuel.

**Keywords**--- Cost of Living, Low Income Household, Malaysia.

## I. Introduction

Generally, there are various factors that cause an increase in the cost of living. One of the factors is that increase in income is not parallel with increase in price of goods and services (Nuradilla Noorazam, 2016). Changes in technology, as well as individual tastes and preferences, could also affect cost of living. It is understandable that with changing time and advancement in technology, items previously considered as luxury might nowadays become necessity and their usage become more difficult to avoid (Zulkifly Omar, 2016).

According to Sinar Harian (2017), the Malaysian public regard the good and services tax (GST) implemented on April 2015 as one of the primary reason for rising cost of living.

Study conducted by Prof Datuk Redzuan Othman involving 4468 respondents from around the country found that 82 percent of respondent agreed that GST has some effect on cost of living and resist its implementation (Sinar Harian, 2017).

Cost of living is highly related to the well-being of society. Based on report, the Malaysian social welfare index improves from year to year, which indicate improvement in the people's overall welfare. For a more specific example, the social welfare index rose from 100 to 121 from year 2000 to 2012 while the Malaysian Well-being Index (MWI)increased from 100 to 133.3 over the same period (Rohami Shafie, 2013). This improvement in Malaysian Well-being index (MWI) and reduction in the poverty rate is consistent with the objective of reaching high income nation by year 2020 (Rohami Shafie, 2013).

Despite the report of reduction in poverty rate and increase in MWI, the issue of rising cost of living continues to become the main concern and one of the main focus of discussion in mainstream media. Accordingly, various effort was done on the part of the government to reduce the burden of rising cost of living on the people.

According to Zulkifly Omar, a senior research fellow of the Malaysian Institute of Economic Research (MIER), the rising cost of living will be felt by individuals when the increase in the prices of goods exceeds the increase in income received. This results in an increase in individual total spending even if there is no alteration in the spending pattern of the individual. In other words, to ensure that their monthly expenses are sufficient, the individual needs to reduce their expenses (Zulkifly Omar, n.d)

This articleis about income and source of income of the head of household in the Kubang Pasu district, their spending behavior and the practices they take to address the rising costof living. The next section explain the methodology, followed by findings and discusson, and finally the conclusion.

## II. Methodology

This study uses primary data obtained using structured questionnaire. The data was collected through face to face interview conducted by trained interviewer with selected respondents. Respondentsare selected among the heads of households (HH) who have monthly income of RM3800 and below. Thus, the interviewer would go from house to house in the villages located in the two districts using the house condition of prospective respondents's as guide. Respondents were selected and interview conducted after it was confirmed that the head of household earn less than RM3800 per month. The data collection process took two months to complete, from October to November 2017.

The interviewers had successfully interviewed 124 respondents. All target respondents are those who are in the lowest 40 percent in the income hierarchy (B 40). In other words, they consist of the head of household (HH) with income below RM3800 per month . This study focuses on area around Mukim Temin and MukimKeplu in Kubang Pasu District. Data is analyzed and presented in the form of descriptive analysis and tables to facilitateclearerunderstanding .

## III. Empirical Results

The following table shows preliminary information which was obtained related to the background of head of household. The table shows that a total of 104, or 83.9 percent, of the household heads are male and the rest, 20 persons (16.1 percent), are women.

Table 1: Respondent Demographic Information

| Item | Frequency | Percentage |
| --- | --- | --- |
| Age | | |
| 20-30 | 4 | 3.2% |
| 31-40 | 17 | 13.7% |
| 41-50 | 32 | 25.8% |
| 51-60 | 41 | 33.1% |
| 61-70 | 20 | 16.1% |
| 71-79 | 8 | 6.5% |
| > 80 | 2 | 1.6% |
| Total | 124 | 100.0% |
| Marital Status | Frequency | Percentage |
| Married | 95 | 76.6% |
| Widow / Widower | 25 | 20.2% |
| Divorced | 4 | 3.2% |
| Total | 124 | 100.0% |
| Education Level | Frequency | Percentage |
| No certificate/education | 7 | 5.6% |
| UPSR or Primary 5 | 8 | 6.5% |
| LCE / PMR / SRP | 36 | 29.0% |
| SPM | 62 | 50.0% |
| STPM / HSC | 3 | 2.4% |
| Diploma | 5 | 4.0% |
| Degree | 1 | 0.8% |
| Master | 2 | 1.6% |
| Total | 124 | 100.0% |

In terms of age, most of the respondent are in the age group between 31 to 70 years, which is 110 persons or 88 percent. Majority, or 76.6 percent, of them are married, while 20.2 percent are widow or widower and only 3.2 percent are divorcee. In terms of education level, only a few, 11 or 8.87 percent, who have education at the STPM level or higher.

In addition to the information on sex, marital status and education level, the study also shows the main types of HH's job. The survey found that most of the head of household involve in agriculture, forestry or fishery sector with 33 persons (26.6 percent). This is expected since the majority of household head in Kedah state are involved in agricultural activities such as rubber tapping or paddy farming, and agriculural income is the most important source of income for HH (Siti Hadijah dan Roslan (2010).

As many as 32 (25.8 percent) of HH involved in services sector, which include working as teacher, running a business, tailoring and others. Out of the 12.1 percent who involves in the transport and storage sector, majority of them working as truck driver.

Table 2: Head of Household Employment Sector

| Employment sector | Frequency | Percentage |
|---|---|---|
| Agriculture, forestry & fisheries | 33 | 26.6% |
| Manufacturing | 6 | 4.8% |
| Electricity, gas, steam & air conditioning | 2 | 1.6% |
| Water supply, sewerage, waste management & restoration activities | 1 | 0.8% |
| Construction | 7 | 5.6% |
| Wholesale & retail | 2 | 1.6% |
| Transport & Storage | 15 | 12.1% |
| Food Stall / Restaurant | 3 | 2.4% |
| Services | 32 | 25.8% |
| Unemployed/Retired | 23 | 18.5% |
| Total | 124 | 100.0% |

## Head of Household Employment and Income

The following table shows the employment sector of respondents and their earnings. Most respondents (26.6 percent) works in agriculture, forestry and fisheries sector, and most of them earn less than RM2000 monthly. 25 percent of respondents works in the services sector, most of whom earned less than RM2000 monthly. Overall, the average respondent received monthly income of less than RM2000 regardless of sector in which they involve.

Table 3: The Relationships between Employment with Income

| Employment sector | Under 1000 | 1001-2000 | 2001-3000 | >3000 | Total |
|---|---|---|---|---|---|
| Agriculture, forestry & fisheries | 19 115.32%) | 11(8.87%) | 3(2.42%) | 0 (0%) | 33(26.61%) |
| Manufacturing | 3(2.42%) | 2(1.61%) | 1(0.81%) | 0(0%) | 6(4.84%) |
| Electricity, gas, steam & air conditioning | 2(1.61%) | 0 (0%) | 0(0%) | 0(0%) | 2(1.61%) |
| Water supply, sewerage, waste management & restoration activities | 1(0.81%) | 0(0%) | 0(0%) | 0(0%) | 1(0.81%) |
| Construction | 5(4.03%) | 2(1.61%) | 0(0%) | 0(0%) | 7(5.65%) |
| Wholesale& Retail | 1(0.81%) | 1(0.81%) | 0(0%) | 0(0%) | 2(1.61%) |
| Transportation &Storage | 8(6.45%) | 4(3.23%) | 2(1.61%) | 1(0.81%) | 15(12.10%) |
| Food Stall, Restaurant | 2(1.61%) | 1(0.81%) | 0(0%) | 0(0%) | 3(2.42%) |
| Services | 18(14.52%) | 9(7.26%) | 5(4.03%) | 0(0%) | 32(25.81%) |
| Unemployed | 15(12.10%) | 6(4.84%) | 1(0.81%) | 1(0.81%) | 23(18.55%) |
| Total | 74 (59.68%) | 36 (29.03%) | 12 (9.68%) | 2 (1.61%) | 124 (100%) |

Table 4: Total Household Income

| Total household income (RM) | Frequency | Percentage |
|---|---|---|
| 0 < 1000 | 44 | 35.5% |
| 1 1 001-2000 | 55 | 44.4% |
| 2 2 001-3000 | 18 | 14.5% |
| 3 > 3 001 | 7 | 5.6% |
| Amount | 124 | 100.0% |

Table 4 shows the total household income which include HH main income, side income, such as rent, income from part time jobs and financial assistance received from the authority such as BR1M, zakat and others. Household income also includes income received by other members of the household including spouses, children, and grandchildren, that is if they live together. This amount is taken into account as not all household heads are employed and have income.

Based on the table, it is shown that 44.4 percent of the total households have income between RM1001 and RM2000. This amount represents 55 households out of a total of 124 households. Followed by 44 households,(35.5 percent),which earns a total income of less than RM1000.

*Household Spending Behavior*

The study also analyzes respondents' spending patterns . The four main items selected to see the pattern of respondents' spending patterns are expenditure on food, utilities,transport and education. There are other items related to spending, however most respondent fail to supply the required information. These other items include insurance, medical expenses and housing costs .

Most respondents do not pay for any insurance policy whether for themselves or their children's education. This is because they have no children in school and they also do not know about the importance of such a policy. The average respondents only visitgovernment health clinic providing free services. They are comfortable to go to the government health centers and clinics 1 Malaysia close to their homes. The majority of respondents live in their own home or country house that do not cost anything to rent.

Table 5: Spending Allocation

| Type of expenditure | Frequency | Percentage |
|---|---|---|
| Food | 104 | 83.9% |
| Utility | 4 | 3.2% |
| Transportation | 14 | 11.3% |
| Education | 2 | 1.6% |
| Total | 124 | 100.0% |

Table 5 shows the behavior of respondents in allocating their income for expenditure. From the total of 124 respondents, 104respondents allocate more income tospend on food. With the majority of them earning less than RM3000, which is more or less on subsistence level, most of the respondents spend most of their money on food items.

A total of 14 respondents (11.3 percent) use most of their income for transportation. 3.2% use for utility expenditures, electricity bill payment, water bills and astro monthly payments, if respondents subscribe to astro sevices, and the remaining 1.6 percent uses most of the income they receive each month for educational expenses.

*Methods of Overcoming Rising Cost of Living*

In dealing with the rising cost of living, the respondents suggest a few method to go around it. Oneisbybeing thrifty and spending according to real needs, taking multiple jobs, kitchen farming, doing overtime work and reducing utility use. If this is done wisely, it can certainly save them the daily or monthly expenses of their family.

There are five steps of limiting spending that are given by the household as shown in Table 6. Among the most common measures taken by households to reduce the effects of rising living cost is tospend only on necessity in order to save (46.8 percent). There are various alternatives to be more thrifty like avoiding to shop on an empty stomach, and preparing shopping list beforehand. If there is no shopping list, we tend to buy more than what is necessary and result in unnecessary spending. Bulk buying to keep for future use and buying during sales and price offer can also be practiced.

In addition, some also suggest the use of creativity, for example instead of spending for normal purchases, spare and leisure times can be used to produced food items that are normally bought like biscuit or cookies, cakes and the likes. Those items that can be consumed, given as gift or sold for income. Other measures also possible, like being more diligent before making purchase such as doing price and quality comparison.

Some of them, 21.88 percent, recourse to doing part time jobs. Working on part time jobs improves their income base to ensure that income is sufficient to cover monthly expenses. Some of those who involves in agriculture, for example, takes part time jobs like taxi driving, doing part time business, tailoring or other entrepreneurial activities.

The non agricultural income is crucial not only to relieve the burden of rising cost of living among farming family, it is also important in reducing poverty and excelerating the economic mobility of low income family (Siti Hadijah et al 2012, Norzita & Siti Hadijah, 2014)  Siti Hadijah et al (2017) also found that income from entrepreneurial activities carried out by a single mother could shorten the length of time to exit from poverty by as much as 25.07 years.

It is also found that some of them chooses to work overtime (9.38 percent). Taking overtime, meaning spending more time on their regular job, it does not involve extra time consuming creativity or extra expenses like tansportation cost. Additionally, 10.94 percent of household heads resort to ideas to reduce cost and spending particularly reducing utility bill by being more observant in their uses.They can conserve by turning off lights and other electrical apppliances when not in use.  Similar observant attitude could be applied in water and smartphones data usage.

Table 6: Methods to deal with rising living cost

| Action | Frequency | Percentage |
|---|---|---|
| Thrift/necessity shopping | 30 | 46.88 % |
| Part time jobs | 14 | 21.88 % |
| Kitchen Farming | 7 | 10.94 % |
| Working overtime | 6 | 9.38 % |
| Reduce Utility bill | 7 | 10.94 % |
| Total | 64 | 100.0% |

Note: Based on first responses of respondents answering this question.

The actions described earlier are possible actions that can be applied on the part of household. But the head of the household also suggested that the government or the responsible party must also contribute to ease the bruden of the increasing cost of living. The study found that as many as 38 respondents (30.65 percent) suggest that the GST be abolished. This means that even though they live in rural areas and most of them are old they also know about current issues that say the GST is causing rise in price of goods and consequently leading to rising cost of living.

The other major suggestion for the the government to control the price of goods and fuel. This control is important so that traders will not simply increase the price of goods. Most households found the price of goods they purchased in retail stores isdifferent from pices in large super markets. But they can not afford the trip to super market for reasons of distance and transportation. A total of 26 respondents (20.97percent) thought that controlling the price might be able to control the rising cost of living. There are also those who suggest that the disbursement of government financial assistance like BRIM to be improved. They argue that with the increase in the amount of BR1M they receive will reduce the impact of rising cost of living. Respondents argue that the existing amount of BR1M is too small compared to their daily expenses.

## IV. Conclusion

Most respondents found to allocate a bigger part of their income for food expenses, followed by transportation, utilities and education. As for covering the extra expenses caused by increasing price, they resort to doing part time jobs, working overtime as well as being more careful in controlling spending. Respondents also propose that the GST to be reexamined. They also asserted the importance of financial assistance from the government in reducing the burden of rising cost on their lives.In conclusion, this study finds that the rising cost of living creates extra pressures and stresses on the public especially the low income group. They have to bear the burden which is ever increasing.Some of them are forced to cut spending in order ensure that their financial standing issufficient to cover their daily sustenance. Additionally, they also need to save and do more than one job to ensure they have enough income every month . Since most of them spend higher propotion of income on food, increased food prices cause their spending to rise. The findings show that financial aid received by them can help ease the burden, especially household head with many dependents. However, the satisfaction level of the received assistance is at moderate level.

## References

[1]     Norzita Jamil & Siti Hadijah Che Mat. (2014). Realiti Kemiskinan Satu Kajian Teoritikal. Jurnal Ekonomi Malaysia. Vol 48 (1) pp 167-177.

[2]     Nuradilla Noorazam. (2016, September 23). Salary is not the main factor of rising cost of living in Malaysia. Astro Awani. Taken on September 23, 2017 from,http://www.astroawani.com/women-bisnes/what-their-factor-the-best-of-the-best-of-sara-hidup-di-malaysia-117473
[3]     Rohami Shafie. (2013, June 27). The balance of the people's economic well-being. Utusan Online. Taking on November 20, 2017, from http://ww1.utusan.com.my/utusan/Plan/20130627/re_07/Welfare- economy-people
[4]     Shaharudin Idrus. (2016, 14 january). Challenging life challenges in big cities.BH Online. Taken on September 21, 2017, from https://www.bharian.com.my/node/114077
[5]     Sinar Online. (2017, March 25). Inflation rises: The lowest income is most affected. Taken on November 19, 2017, from http://www.sinarharian.com.my/mobile/nasional/inflation-the-best-best-results-forest-1.649353
[6]     Siti Hadijah Che Mat and Roslan Abdul Hakim. (2010) Kesan pendapatan bukan pertanian ke atas krmidkinsn: kajian kes petani di Kedah. Jurnal Ekonomi Malaysia. Vol 44. pp101-107
[7]     Siti Hadijah Che Mat, Nor'Aznin Abu Bakar and Ahmad Zafarullah Abdul Jalil. (2012). Pendapatan bukan pertanian dan sumbangannya dalam memendekkan tempoh masa keluar daripada kepompong kemiskinan. Kajian Malaysia. Vol. 30, No. 2. pp 121-141.
[8]     Siti Hadijah Che Mat, Mukaramah Harun, Zalina Zainal and Zakiyah Jamaluddin. (2017). Entrepreneurial Activity and the Time Exit Poverty among Single Mother in Malaysia. International Journal of Applied Business and Economic Research. Vol 15. No 24. Pp 319-330.
[9]     Movasagh, F., Babaei, M., Mohammadi, F., Mirzaee, M.S. The underlying causes of chronic renal failure in children and the elderly (2019) International Journal of Pharmaceutical Research, 11 (1), pp. 1758-1760.
[10]    Zulkifly Omar. (n.d). Addressing the Improvement of Cost of Living. Taken in September 25, 2017 from https://www.mier.org.my/newsarticles/archives/pdf/DrZul21_12_2015.pdf